\documentstyle[11pt]{article}
\begin{document}
{\setlength{\oddsidemargin}{1.2in}
\setlength{\evensidemargin}{1.2in} } \baselineskip 0.55cm
\begin{center}
{ {\bf GUP effects on Hawking temperature in Riemann space-time.}}
\end{center}
\begin{center}
${\rm Y.\ Kenedy \ Meitei}^{1}$. $ {\rm T.\ Ibungochouba\ Singh}^{1}$. ${\rm I.\ Ablu\ Meitei}^{2}$.
\end{center}
\begin{center}
1\ Department of Mathematics, Manipur University, Canchipur, Manipur 795003, India\\
2\ Department of Physics, Modern College, Imphal, Manipur 795005, India\\
E-mail:yumkendy@gmail.com
\end{center}
\date{}

\begin{abstract}
In this paper, the modified Hawking temperature of a static Riemann space-time is studied using the generalized Klein-Gordon equation and the generalized Dirac equation. Applying the Kerner-Mann quantum tunneling method, the modified Hawking temperature for scalar particle and fermions that crosses the event horizon of the black hole have been derived. We observe that the quantum gravity effect reduces the rise of thermal radiation temperature of the black hole.

\end{abstract}


\section{\bf Introduction}  \setcounter {equation}{0}
\renewcommand {\theequation}{1.\arabic{equation}}

The thermodynamics of black holes has been constructed successfully [1] since the discovery of Hawking radiation using quantum field theory in curved space-time [2, 3]. The relationship of the entropy of black hole and its horizon area has been established in [4]. Since then, different authors proposed different methods for studing the Hawking radiation. Damour and Ruffini [5] and  Sannan [6] studied the Hawking radiation using tortoise coordinate transformation. Chandrasekhar [7] and  Bonner and Vaidya [8] have shown that  the Dirac equation and Maxwell's electromagnetic equations can be separated for stationary space-time. However, using tortoise transformation, the Dirac equation and Maxwell's electromagnetic field equations can be separated for a stationary and nonstationary black hole. Following this method, many works have been done in [9-14].

Parikh and Wilczek [15] studied the Hawking radiation as a quantum tunneling process and this method is known as a null geodesic method. The Hawking radiation as a tunneling of particles was also studied using Hamilton-Jacobi method [16]. When outgoing particles tunneled the barrier, the imaginary part of the action can be derived by applying Feynman prescription and WKB approximation. Refs. [17, 18] investigated the Hawking radiation in more complicated black holes by applying the Hamilton-Jacobi method. They showed that the spectrum was no longer thermal.

Kerner and Mann [19] investigated the tunneling of Dirac particle through event horizon for the Rindler black hole and the general rotating black hole. In this method, appropriate Gamma matrices were chosen and wave functions were inserted into the Dirac equation, the action which is related to Boltzmann factor of emission at Hawking temperature according to semiclassical WKB approximate can be obtained. Using this method, the Hawking radiation in different complicated black holes can be achieved  [20-23]. Kruglov [24-25] proposed the thermal radiation from a black hole by using Hamilton-Jacobi ansatz to Proca equation, WKB approximation and Feynman prescription. For the Rindler black hole, the emission temperature is in agreement with the Unruh temperature and for the Schwarzschild space-time, the emission temperature coincides with the Hawking temperature of scalar particle. Applying this method, tunneling of vector particle in different complicated black holes were obtained in [26-29].

The existence of a minimal length [30-34] was shown by quantum gravity such as string theory, loop quantum gravity and quantum geometry. This minimal length can be achieved by using the generalized uncertainty principle (GUP) through modified commutation relation. Modifying the fundamental commutation relation [35-36] $[x_{i}, p_{i}]=i\hbar\delta_{ij}[1+\beta p^2]$, the inequality of GUP is obtained as $\Delta x\Delta p\geq \hbar/2[1+\beta (\triangle p)^2]$, where $\beta=\beta_{0}/M^2_{p}$. $M_{p}$ and $\beta<10^{34}$ are the Plank mass and dimensionless parameter respectively. The position, $x_{i}$ and momentum, $p_{i}$, satisfying the standard commutation relation
$[x_{0i}, p_{0i}]=i\hbar\delta_{ij},$  can be defined as $x_{i}=x_{0i}$ and $p_{i}=p_{0i}(1+p^{2}_{0i})$ respectively. Das et al. [37-38] investigated the GUP based on doubly special relativity. The Unruh effect has also been studied by Majhi and Vagenas [39] based on a modified form of GUP. The radiation of massless scalar field in the Schwarzschild black hole has been investigated in [40] by taking quantum gravity into account influenced by DSR-GUP and Parikh and Wilczek tunneling method. It is observed that the remnant of black hole evaporation is $\geq\frac{M_{p}}{\beta}$. Following their works, many interesting results have been derived in [41-51]. Recently, Ablu, et al [52] discussed the tunneling of scalar particle for BTZ black hole using the generalized Klein-Gordon equation based on GUP and the entropy correction at the black hole event horizon has been recovered.

The aim of this paper is to investigate the correction of Hawking temperature of scalar particles and fermions crossing the black hole horizon of a Riemann space-time by taking quantum gravity into account. Applying generalized Klein-Gordon equation and generalized Dirac equation based on GUP, the corrected Hawking temperature has been recovered.

The paper is organized as follows. In section 2, the Klein-Gordon equation and the Dirac equation based on GUP have been revisited. In section 3, the correction of Hawking temperature of Reimann space-time is investigated by applying generalized Klein-Gordon equation and WKB approximation. In section 4, the tunneling of fermions across in the Reimann space-time is investigated by using generalized Dirac equation and WKB approximation. In section 5, some conclusions are given.

\section{\bf Revisit of Generalized Klein-Gordon equation and Dirac equation }  
\renewcommand {\theequation}{\arabic{equation}}

The Klein-Gordon equation of scalar particle having mass $m_0$ in four dimensional space without an electromagnetic field is given by
\begin{eqnarray}
-p^{i}p_{i} = m_{0}^{2}.
\end{eqnarray}

 To study quantum gravity effect, the above field equation can be written as

\begin{eqnarray}
-(i\hbar)^{2}\partial^{t}\partial_{t} = (i\hbar)^{2}\partial^{i}\partial_{i} + m_{0}^{2}.
\end{eqnarray}
The modified expression of energy in quantum theory of gravity can be expressed as [53, 54]
\begin{eqnarray}
\tilde{E} = E(1-\beta E^2) = E[1-\beta(p^2 + m_{0}^{2})],
\end{eqnarray}
where  $E=i\hbar\partial_{t}$ and $E^2-p^2 = m_{0}^{2}$ are the energy operator and energy mass shell respectively. Using modified momentum operators and the dispersion relation into the above equation, the generalized Klein-Gordon equation [46] can be written as
\begin{eqnarray}
-(i\hbar)^{2}\partial^{t}\partial_{t}\psi = \{(-i\hbar)^{2}\partial^{i}\partial_{i} + m_{0}^{2}\}[1-2\beta\{(-i\hbar)^{2}\partial^{i}\partial_{i} + m_{0}^{2}\}]\psi.
\end{eqnarray}
The Dirac equation in four dimensional curved space-time is given by [55]
\begin{eqnarray}
i\gamma^{a}(\partial_a+\Omega_a)\psi+\frac{m}{\hbar}\psi=0, \,\,   \Omega_a=\frac{i}{2}\omega^{bc}_{a}\, \Sigma_{bc},
\end{eqnarray}
where $\omega^{bc}_{a}$ is the spin coefficients and $\Sigma^{,s}_{bc}$ will satisfy the following conditions

\begin{eqnarray}
\Sigma_{bc}=\frac{i}{4}[\gamma^{b},  \gamma^{c}],  \,\,\,\,\{\gamma^{b},\,\,\gamma^{c}\}=2\eta^{bc}.
\end{eqnarray}
The square of momentum operator is
\begin{eqnarray}
p^{2} = p_{i}p^{i} \simeq -\hbar^{2}[\partial^{i}\partial_{i}-2\beta\hbar^{2}(\partial^{i}\partial_{i})(\partial^{i}\partial_{i})].
\end{eqnarray}

To obtain generalized Dirac equation based on GUP in curved space-time, Eq. (5) can be written as
\begin{eqnarray}
-i\gamma^0\partial_0\psi = (i\gamma^i\partial_i + i \gamma^a\Omega_a + \frac{m_{0}}{\hbar})\psi,
\end{eqnarray}
here $i = 1, 2, 3$ indicates the spatial coordinates. Using Eqs. (3) and (7) in Eq. (8), we obtain generalized form of Dirac equation as,
\begin{eqnarray}
&& [i\gamma^0\partial_0 + i\gamma^i(1 - \beta{m_0}^2)\partial_i + i\gamma^i\beta\hbar^2(\partial_j\partial^j)\partial_i + \frac{m_0}{\hbar}(1-\beta{m_0}^2 + \beta\hbar^2\partial_j\partial^j)\cr
&& + i\gamma^{\mu}\Omega_{\mu}(1- \beta m_{0}^{2} + \beta\hbar^{2}\partial_j\partial^j)]\psi = 0,
\end{eqnarray}
where $\psi$ is a Dirac spinner wave function.

\section{\bf Hawking temperature of Riemann space-time for scalar particle}. 
\renewcommand {\theequation}{\arabic{equation}}

 The line element of static Riemann space-time in four dimensional space-time $(t, x, y, z)$ can be written [56] as
\begin{eqnarray}
ds^2&=&-a^2dt^2 + b^2dx^2 + c^2dy^2 + d^2dz^2,
\end{eqnarray}
where $a, b, c$ and $d$ are the functions of $(x,y,z)$. Eq. (10) has an event horizon at $x = \xi$.
According to Ref. [57], the contravariant and covariant components of Riemann space-time can be written as
\begin{eqnarray}
  g_{00} = -q^2(x - \xi) = -a^2,\, g^{11} = p^2(x,y,z)(x - \xi) = \frac{1}{b^2}, \, g^{22} = \theta,\, g^{33} = \varphi.
\end{eqnarray}
 The position of event horizon is $x = \xi$, and $q^2, p^2, \theta, \varphi$ are arbitrary non-zero and non-singular functions at the event horizon.
The surface gravity $\kappa$ is [57]
\begin{eqnarray}
  \kappa =\lim_{g_{00}\rightarrow 0}\frac{1}{2}\sqrt{-\frac{g^{11}}{g_{00}}}\frac{\partial g_{00}}{\partial x} = \frac{1}{2}p(\xi)q(\xi)
\end{eqnarray}
and the temperature of the black hole is given by
\begin{eqnarray}
  T_0 =\frac{p(\xi)q(\xi)}{4\pi}.
\end{eqnarray}

Substituting covariant and contravariant components of Eq. (10) into generalized Klein-Gordon equation given in Eq. (4), we have

\begin{eqnarray}
  && -\frac{\hbar^2}{a^2}\frac{\partial^2 \psi}{\partial t^2} = -\hbar^2(\frac{1}{b^2}\frac{\partial^2\psi}{\partial x^2} + \frac{1}{c^2}\frac{\partial^2\psi}{\partial y^2} + \frac{1}{d^2}\frac{\partial^2\psi}{\partial\ z^2})
    - 2\beta \hbar^4(\frac{1}{b^2}\frac{\partial^2}{\partial x^2}\cr&& + \frac{1}{c^2}\frac{\partial^2}{\partial y^2} + \frac{1}{d^2}\frac{\partial^2}{\partial z^2})(\frac{1}{b^2}\frac{\partial^2\psi}{\partial x^2} + \frac{1}{c^2}\frac{\partial^2\psi}{\partial y^2} + \frac{1}{d^2}\frac{\partial^2\psi}{\partial z^2}) \cr&&+ 4\beta {m_0}^2\hbar^2(\frac{1}{b^2}\frac{\partial^2\psi}{\partial x^2} + \frac{1}{c^2}\frac{\partial^2\psi}{\partial y^2} + \frac{1}{d^2}\frac{\partial^2\psi}{\partial z^2}) + {m_0}^2(1 - 2\beta {m_0}^2)\psi.
\end{eqnarray}
To investigate the correction of Hawking temperature of Riemann space-time based on the GUP, the wave function is chosen as
\begin{eqnarray}
\psi = Ae^{\frac{i}{\hbar}S(t,x,y,z)}.
\end{eqnarray}
Using Eq. (15) in Eq. (14), the following relation is obtained as
\begin{eqnarray}
&&-\frac{1}{a^2}(\frac{\partial S}{\partial t})^2 = \Big[ \frac{1}{b^2}(\frac{\partial S}{\partial x})^2 + \frac{1}{c^2}(\frac{\partial S}{\partial y})^2 + \frac{1}{d^2}(\frac{\partial S}{\partial z})^2 + {m_0}^2\Big] \cr&&\times \Big[ 1 - 2\beta\Big\{\frac{1}{b^2}(\frac{\partial S}{\partial x})^2 + \frac{1}{c^2}(\frac{\partial S}{\partial y})^2 + \frac{1}{d^2}(\frac{\partial S}{\partial z})^2 + {m_0}^2 \Big\}\Big].
\end{eqnarray}
Using the separation of variables of the form
\begin{eqnarray}
S = -\omega t + R(x) + W(y,z),
\end{eqnarray}
where $\omega$ denotes the energy of the emitted scalar particle.
We know that the Hawking radiation takes place along the radial direction, then we take
\begin{eqnarray}
 \frac{1}{c^2}(\frac{\partial S}{\partial y})^2 + \frac{1}{d^2}(\frac{\partial S}{\partial z})^2 = u.
\end{eqnarray}
The value of $u$ is taken as a constant and can be put zero. Using Eqs. (17) and (18) into Eq. (16), a biquadratic equation is obtained as follows
\begin{eqnarray}
A(R^{'}_x)^4 + B{(R^{'}_x)}^2 + C = 0,
\end{eqnarray}
where
\begin{eqnarray}
A = -\frac{2\beta}{b^4}, \;\;\; B = (1 - 4\beta m_0^2)\frac{1}{b^2}, \;\;\; C = m_0^2(1 - 2\beta m_0^2) - \frac{\omega^2}{a^2}.
\end{eqnarray}
Eq. (19) has four roots of which only two roots have physical meaning and these are given by
\begin{equation}
  R_{\pm} = \pm\int\frac{b}{a}\sqrt{\omega^2 - m_0^2a^2 + 2\beta m_0^4a^2}(1 + 2\beta m_0^2)dx,
\end{equation}
where $R_{+} $ indicates the scalar particle moving away from the black hole and $R_{-}$ corresponds to scalar particle approaching toward the black hole. Using Eq. (11) and completing the integral of Eq. (21), the imaginary part of radiant action is given by
\begin{eqnarray}
\textrm{Im}R_{\pm}={\pm}\frac{2\pi\omega}{p(\xi)q(\xi)}(1 + 2\beta m_0^2).
\end{eqnarray}
The tunneling probability of the scalar particle that has crossed the black hole event horizon  is
\begin{eqnarray}
\Gamma=\frac{\rm{Prob(emission)}}{\rm{Prob(absorption)}}
&=& \frac{\exp(-\textrm{Im}R_{+} -\textrm{Im}W)}{\exp(-\textrm{Im}R_{-} -\textrm{Im}W)}\cr
&=& \exp\Big({\frac{4\pi\omega}{p(\xi)q(\xi)}(1 + 2\beta m_0^2)}\Big).
\end{eqnarray}
The corrected Hawking temperature is given by
\begin{eqnarray}
T= \frac{p(\xi)q(\xi)}{4\pi(1 + 2\beta m_0^2)}=\acute{T}_0(1 -2\beta m_0^2),
\end{eqnarray}
where $\acute{T}_0 = \frac{p(\xi)q(\xi)}{4\pi}$ is the actual thermal radiation temperature of Riemann space-time. Thus, the correction to the Hawking temperature due to quantum gravity effects  has been obtained. It is observed that the correction to the Hawking temperature of Riemann space-time depends on the mass of the emitted scalar particle. From Eqs. (13) and (24), it is observed that the quantum gravity effects can lower the rise of Hawking temperature of Riemann space-time. When $\beta = 0$, the original Hawking temperature is recovered. If we ignore a small term, $\frac{2\beta m_{0}^{4}a^{2}}{\omega^{2}}$ in the roots of Eq. (19), then 

\begin{eqnarray}
  R_{\pm} &=& \pm\int\frac{b}{a}\sqrt{\omega^2 - m_0^2a^2 + 2\beta m_0^4a^2}[1 + \beta (m_0^2 + \frac{\omega^{2}}{a^{2}})]dx,\cr
          &=& \pm\frac{2\pi i \omega}{p(\xi)q(\xi)}[1 + \beta\{\frac{3m_{0}^{2}}{2} - \omega^{2}(\frac{p(\xi)q'(\xi) + q(\xi)p'(\xi)}{p(\xi)q^{3}(\xi)} )\}].
\end{eqnarray}
where $p'(\xi) = \frac{\partial p}{\partial x}|_{x = \xi} $ and $q'(\xi) = \frac{\partial q}{\partial x}|_{x = \xi}$. The corrected Hawking temperature of Riemann space-time is
\begin{equation}
T = \frac{p(\xi)q(\xi)}{4\pi}[1 - \beta \chi]
\end{equation}
where $\chi = \frac{3m_{0}^{2}}{2} - \omega^{2}(\frac{p(\xi)q'(\xi) + q(\xi)p'(\xi)}{p(\xi)q^{3}(\xi)} )$. In this case, the quantum gravity effects lower the rise of Hawking temperature in Riemann space-time and the Hawking temperature depends on the mass and energy of the emitted particle and also on arbitrary non-zero and non-singular functions

\section{\bf Dirac Equation:}

To investigate the fermion tunneling across the event horizon of Riemann space-time, our aim is to find the imaginary part of the radiant action. For the Riemann space-time, $\gamma^{a}$ matrices in $(t, x, y, z)$ coordinates system are chosen as

\begin{eqnarray}
\gamma^t&=& \frac{1}{\sqrt{F(r,\theta)}}
\left({\begin{array}{c c c c}
i & 0 & 0 & 0\\
0 & i & 0 & 0\\
0 & 0 & -i & 0\\
0 & 0 & 0 &-i\\
\end{array}}\right),
\gamma^x = \sqrt{G(r, \theta)} \left({\begin{array}{c c c c}
0 & 0 & 1 & 0\\
0 & 0 & 0 &-1\\
1 & 0 & 0 & 0\\
0 & -1 & 0 & 0\\
\end{array}}\right),\cr
\gamma^y &=& \frac{1}{K(r, \theta)}\left({\begin{array}{c c c c}
0 & 0 & 0 & 1\\
0 & 0 & 1 & 0\\
0 & 1 & 0 & 0\\
1 & 0 & 0 & 0\\
\end{array}}\right),
\;\;\;\gamma^y = \frac{1}{H(r, \theta)}\left({\begin{array}{c c c c}
0 & 0 & 0 & -i\\
0 & 0 & i & 0\\
0 & -i & 0 & 0\\
i & 0 & 0 & 0\\
\end{array}}\right).
\end{eqnarray}
The tunneling of Dirac particle from the Riemann space-time can be investigated by taking the modified wave function as
\begin{eqnarray}
 \psi = exp(\frac{i}{\hbar}S(t,x,y,z))
 \left(
{\begin{array}{c}
   A(t,x,y,z) \\
   0\\
   B(t,x,y,z)\\
   0
 \end{array}}\right),
  \end{eqnarray}
  where $A(t,x,y,z)$ and $B(t,x,y,z)$ are arbitrary functions. Coordinate $S(t,x,y,z)$ is the action of the radiant particle. Substituting Eqs. (27) and (28) in Eq. (9) and neglecting the first order term of $\hbar$, we get the following four equations
 \begin{eqnarray}
&& [-\frac{i}{\sqrt{F(r, \theta)}}\frac{\partial I}{\partial t} + m_{0}(1 - \beta m_{0}^{2}) - m_{0}\beta\{G(r, \theta)(\frac{\partial I}{\partial x})^2 + \frac{1}{K^{2}(r, \theta)}(\frac{\partial I}{\partial y})^2 + \frac{1}{H^2(r, \theta)}(\frac{\partial I}{\partial z})^2\} ]A\cr
&& + \sqrt{G(r, \theta)}[\frac{\beta}{b^{2}}(\frac{\partial I}{\partial x})^2 + \beta\{\frac{1}{c^2}(\frac{\partial I}{\partial y})^2 + \frac{1}{d^2}(\frac{\partial I}{\partial z})^2\} - (1 -\beta m_{0}^{2})](\frac{\partial I}{\partial x}) B = 0.\\
&&\frac{1}{b}[\beta\{\frac{1}{b^{2}}(\frac{\partial I}{\partial x})^2 + \frac{1}{c^2}(\frac{\partial I}{\partial y})^2 + \frac{1}{d^2}(\frac{\partial I}{\partial z})^2\} - (1 -\beta m_{0}^{2})](\frac{\partial I}{\partial x})A\cr
&& + [\frac{i}{a}\frac{\partial I}{\partial t} + m_{0}(1 - \beta m_{0}^{2}) - m_{0}\beta\{\frac{1}{b^{2}}(\frac{\partial I}{\partial x})^2 + \frac{1}{c^{2}}(\frac{\partial I}{\partial y})^2 + \frac{1}{d^2}(\frac{\partial I}{\partial z})^2\} ]B = 0. \\
&& [\frac{1}{c}\frac{\partial I}{\partial y}\{ \beta(\frac{1}{b^{2}}(\frac{\partial I}{\partial x})^2 + \frac{1}{c^2}(\frac{\partial I}{\partial y})^2 + \frac{1}{d^2}(\frac{\partial I}{\partial z})^2)-(1 - \beta m_{0}^{2})\}\cr
&& +\frac{i}{d}\frac{\partial I}{\partial z}\{\beta(\frac{1}{b^{2}}(\frac{\partial I}{\partial x})^2 + \frac{1}{c^2}(\frac{\partial I}{\partial y})^2 + \frac{1}{d^2}(\frac{\partial I}{\partial z})^2)-(1 - \beta m_{0}^{2})\}]A = 0.\\
&& [\frac{1}{c}\frac{\partial I}{\partial y}\{ \beta(\frac{1}{b^{2}}(\frac{\partial I}{\partial x})^2 + \frac{1}{c^2}(\frac{\partial I}{\partial y})^2 + \frac{1}{d^2}(\frac{\partial I}{\partial z})^2)-(1 - \beta m_{0}^{2})\}\cr
&& +\frac{i}{d}\frac{\partial I}{\partial z}\{\beta(\frac{1}{b^{2}}(\frac{\partial I}{\partial x})^2 + \frac{1}{c^2}(\frac{\partial I}{\partial y})^2 + \frac{1}{d^2}(\frac{\partial I}{\partial z})^2)-(1 - \beta m_{0}^{2})\}]B = 0.
\end{eqnarray}
The Riemann space-time has a time like Killing vector $\frac{\partial}{\partial t}$. The separation of variables of Eqs. (29-32) would be difficult, because the radiant action $I$ is a functions of $t, x, y$ and $z$ . In order to separate the variables, the action $I$ can be expressed as
\begin{eqnarray}
I = -\omega t + Z(x) + W(y, z),
\end{eqnarray}
where $\omega$ is the energy of the emitted fermion. Eliminating $A$ and $B$ from Eqs. (31) and (32), the identical equations can be obtained as
\begin{eqnarray}
\Big(\frac{1}{c}\frac{\partial W}{\partial y} +\frac{i}{d}\frac{\partial W}{\partial z}\Big)\Big[\beta\Big\{\frac{1}{b^{2}}\Big(\frac{\partial Z}{\partial x}\Big)^2 + \frac{1}{c^2}\Big(\frac{\partial W}{\partial y}\Big)^2 + \frac{1}{d^2}\Big(\frac{\partial W}{\partial z}\Big)^2\Big\}-\Big(1 - \beta m_{0}^{2}\Big)\Big] = 0.
\end{eqnarray}
From Eq. (34), the second factor inside the square brackets will not be equal to zero. Then we have,
\begin{eqnarray}
\frac{1}{c}\frac{\partial W}{\partial y} +\frac{i}{d}\frac{\partial W}{\partial z} = 0.
\end{eqnarray}
The solution of the above equation is a complex function of $W$. It can be neglected because this solution does not yield any contribution to the tunneling rate. Then we can take $c^{-2}\frac{\partial W}{\partial y} + d^{-2}\frac{\partial W}{\partial z} = 0$. Next, we have to solve the Eqs. (29) and (30) by using Eqs. (33) and (35) to obtain the Hawking radiation of a Riemann space-time at the event horizon. The non-trivial solution of Eqs. (29) and (30) would be obtained only when the determinant of co-efficient matrix of $A(t,x,y,z)$ and $B(t,x,y,z)$ is equal to zero and neglecting the higher-order terms of $\beta$, a biquadratic equation is obtained as:
\begin{eqnarray}
\frac{2\beta}{b^4}(Z^{'}(x))^{4} - \frac{1}{b^2} (Z^{'}(x))^{2} + \{\frac{\omega^{2}}{a^{2}} + m_{0}^{2}(1 + 2\beta m_{0}^{2})\} = 0.
\end{eqnarray}
The required two roots having physical meaning of the above equation are given by
\begin{eqnarray}
Z(x)_{\pm} &=& \pm \int \frac{b}{a}\sqrt{(\omega^{2} + m_{0}^{2}a^{2})} [1 + \beta(m_{0}^{2} + \frac{\omega^{2}}{a^{2}})]dx\cr
     &=& \pm\frac{2\pi i \omega}{p(\xi)q(\xi)}[1 + \beta\{\frac{3m_{0}^{2}}{2} - \omega^{2}(\frac{3p(\xi)q'(\xi) + q(\xi)p'(\xi)}{p(\xi)q^{3}(\xi)} )\}].
\end{eqnarray}
The tunneling probability of the fermion crossing the event horizon is
 \begin{eqnarray}
 \Gamma = \exp(\frac{4\pi\omega}{p(\xi)q(\xi)}(1 + \beta\Pi)),
 \end{eqnarray}
where $\Pi = \frac{3m_{0}^{2}}{2} - \omega^{2}(\frac{3p(\xi)q'(\xi) + q(\xi)p'(\xi)}{p(\xi)q^{3}(\xi)} )$.
The corrected Hawking temperature is given by
\begin{eqnarray}
 T = \frac{p(\xi)q(\xi)}{4\pi}(1 - \beta\Pi)= T_{0}(1 - \beta\Pi),
 \end{eqnarray}
 where $T_{0}= \frac{p(\xi)q(\xi)}{4\pi} $ is the standard Hawking temperature of the Riemann space-time. From the above equation, we can conclude that the corrected Hawking temperature is not only related to the mass of black hole but also the energy and mass of the emitted particle. If $\beta = 0$, the corrected Hawking temperature becomes standard Hawking temperature.

\section{\bf Conclusions} \setcounter {equation}{0}

We have investigated the tunneling of scalar particle and fermion crossing the horizon of a static Riemann space-time by using the generalized Klein-Gordon equation and the generalized Dirac equation respectively. For the tunneling of a scalar particle, the actual calculation shows that the modified Hawking temperature depends on the mass of the emitted particle. If we ignore a small term $\frac{2\beta m_{0}^{4}a^{2}}{\omega^{2}}$ in the roots of a biquadratic equation having physical meaning, the modified Hawking temperature of a Riemann space-time is found to depend not only on  mass of the emitted particle but also on the energy of the emitted particle. For the fermions tunneling the event horizon of a Riemann space-time, the actual calculation indicates that the modified Hawking temperature of a Riemann space-time depends not only on mass of the emitted particle, but also on the energy of the emitted particle.

For fermions tunneling, it is observed that
\begin{itemize}
 \item If $\frac{3m_{0}^{2}}{2\omega^{2}}  = \frac{3p(\xi)q'(\xi) + q(\xi)p'(\xi)}{p(\xi)q^{3}(\xi)}$, the effect of GUP has been cancelled and standard Hawking temperature of Riemann space-time is recovered.
 \item If $\frac{3m_{0}^{2}}{2\omega^{2}} > \frac{3p(\xi)q'(\xi) + q(\xi)p'(\xi)}{p(\xi)q^{3}(\xi)}$, the effect of GUP will reduce the rise of Hawking temperature of Riemann space-time.
 \item  Lastly if $\frac{3m_{0}^{2}}{2\omega^{2}}  < \frac{3p(\xi)q'(\xi) + q(\xi)p'(\xi)}{p(\xi)q^{3}(\xi)}$, the effect of GUP will rise the Hawking temperature of Riemann space-time.
\end{itemize}
Similar conclusion can be drawn for the scalar particle. It is worth mentioning that the presence of GUP will reduce the rise of Hawking temperature in black holes.

{\bf Acknowledgement}s : The author YKM acknowledges CSIR for providing financial support.

\end{document}